\documentclass[aps,pra,twocolumn,groupedaddress,showpacs]{revtex4}

\usepackage{graphicx}
\usepackage{amsmath}
\usepackage{bm}
\usepackage{color}

\newcommand{\Eq}[1]{Eq.~(\ref{#1})}
\newcommand{\Figure}[1]{Figure~\ref{#1}}
\newcommand{\Fig}[1]{Fig.~\ref{#1}}
\def\beq{\begin{equation}} \def\eeq{\end{equation}}
\def\bea{\begin{eqnarray}} \def\eea{\end{eqnarray}}
\def\bse{\begin{subequations}} \def\ese{\end{subequations}}

\def\||{\parallel}

\def\<{\left\langle} \def\>{\right\rangle}
\def\({\left(} \def\){\right)}
\def\[[{\left[} \def\]]{\right]}

\begin{document}


\title{Phase-ordering percolation and an infinite domain wall\\ in segregating binary Bose-Einstein condensates}


\author{Hiromitsu Takeuchi}
\email{hirotake@sci.osaka-cu.ac.jp}
\homepage{http://hiromitsu-takeuchi.appspot.com/}
\author{Yumiko Mizuno}
\author{Kentaro Dehara}
\affiliation{Department of Physics, Osaka City University, 3-3-138 Sugimoto, Sumiyoshi-ku, Osaka 558-8585, Japan}


\date{\today}

\begin{abstract}
Percolation theory is applied to the phase-transition dynamics of domain pattern formation in segregating binary Bose--Einstein condensates in quasi-two-dimensional systems.
Our finite-size-scaling analysis shows that the percolation threshold of the initial domain pattern emerging from the dynamic instability is close to 0.5 for strongly repulsive condensates.
The percolation probability is universally described with a scaling function when the probability is rescaled by the characteristic domain size in the dynamic scaling regime of the phase-ordering kinetics, independent of the intercomponent interaction.
It is revealed that an infinite domain wall sandwiched between percolating domains in the two condensates has an noninteger fractal dimension and keeps the scaling behavior during the dynamic scaling regime.
 This result seems to be in contrast to the argument that the dynamic scale invariance is violated in the presence of an infinite topological defect in numerical cosmology.
\end{abstract}


\maketitle

\section{INTRODUCTION}\label{Intro}
The phase-transition dynamics of spontaneous symmetry breaking (SSB) is an important problem discussed in different fields such as condensed matter physics, cosmology, and high-energy physics \cite{1995Vilenkin,2002Onuki,2006Vachaspati,2000Bunkov}. It is believed that the SSB transition causes the nucleation of topological defects due to the Kibble--Zurek mechanism (KZM) in the early universe \cite{1976Kibble,1996Zurek}. Because it is difficult to examine the KZM in our universe experimentally, the defect-nucleation problem has been attracting much attention in the systems of helium superfluids and atomic Bose--Einstein condensates (BECs), given the analogy in SSB between our universe and the quantum fluids \cite{2000Bunkov,1996Zurek,2003Volovik,1994Hendry, 1996Bauerle,1996Ruutu,Sadler2006,2008Weiler}.

Linear and planar defects such as strings and domain walls form a complicated network in quenched phase-transition dynamics. The nucleated defect is destroyed if it is closed; e.g., a closed loop of a quantized vortex (cosmic string) can shrink and finally collapse owing to the interactions with phonons and/or quasiparticles (gravitational waves) \cite{1964Rayfield,1991Kopnin,2001Berezinsky,2005Damour}, except for a specific case known as vortons \cite{1995Vilenkin,2003Volovik,2004Metlitski}. An infinite defect, an open defect across the system, can survive. It is important to study infinite defects because they could survive for a long time after the SSB phase transition begins and thus can exert a dominant influence on the later dynamics. In addition, the solution to this problem can provide an answer to the interesting question of whether or not infinite defects are present in our current universe.

In cosmology, it is suggested that percolation is important for understanding the defect-survival problem \cite{1984Vachaspati,2000Vachaspati}. Standard percolation (SP) theory \cite{1994Stauffer} aims to describe the quite common problem of how a connected element ``percolates'' (penetrates through the system). The critical scaling behavior appears when the occupation rate $p$ of the element considered approaches a threshold $p_c$ over which there appears a percolating cluster.
The critical behavior is characterized by a set of universal critical exponents,
 which describe e.g. the fractal behavior of the percolating cluster and the size distribution of non-percolating ones independent of microscopic details of the physical systems.
Regarding a real scalar field as the order parameter due to SSB under consideration, a domain wall is formed along the boundary between two spatial domains with positive and negative values in the field. Therefore, an infinite domain wall can appear when positive and negative domains simultaneously percolate with $p_c \leq 0.5$ since the simultaneous percolation does not occur with $p_c>0.5$.
The study of infinite domain walls is fundamental to understanding infinite strings because an infinite string may be considered as the intersection of two infinite domain walls in the real and imaginary parts of a complex scalar field \cite{1994Stauffer}. These arguments sound quite reasonable, and it is meaningful to tackle the defect-survival problem in such a direction by utilizing the quantum fluid systems.
However, it seems difficult to treat the problem in later stages of phase-transition dynamics because of its complicated non-equilibrium development of order parameters.

An empirical law based on the dynamic scaling hypothesis in phase-ordering kinetics \cite{Bray1994} sheds light on this problem.
This law states that spatial patterns of order parameters in earlier stages of phase-transition dynamics are statistically similar to those in later stages if the patterns are scaled by the characteristic domain sizes.
Since the theoretical analysis in SP theory is directly related to the statistical property of spatial distributions of domains (clusters),
the phase-ordering kinetics can be reinterpreted in terms of percolation problem.
For the sake of convenience, let us refer to such a study as {\it phase-ordering percolation} in this paper.

Within recent years Arenzon {\it et al.} \cite{Arenzon2007} studied the statistics of the areas enclosed by domain boundaries during the curvature-driven coarsening dynamics of a two-dimensional nonconserved scalar field by carrying out numerical simulations on the square-lattice ferromagnetic Ising model from a disordered initial state with $p=0.5$.
They showed that the time development of the distribution of domain  areas is described well by using a dynamic scaling function with a critical exponent of SP theory for $p_c=0.5$. 
Additionally, it is surprising that results of SP theory can predict the final state of the Ising system;
the existence probability of a percolating domain with a specified topology is related to the possibility for the system to develop into the metastable stripe state with open domain walls in the last stage of the relaxation dynamics \cite{Barros2009}.

Although the occupation rate $p$ is not a conserved quantity in those studies \cite{Arenzon2007,Barros2009},
$p$ is kept constant statistically during the time development due to the dynamic scaling law.
Thereby, in the dynamic scaling regime where the system is described by the law,
 we expect that such problems of phase-ordering percolation would be applicable in a similar manner to systems of a conserved scalar field, in which $p$ is conserved exactly.
Study of phase-ordering percolation for both conserved and nonconserved fields develops our understanding of SSB transition dynamics to a higher level.
However, the phase-ordering percolation for a conserved field is not well studied.

In this paper, we theoretically study phase-ordering percolation for a conserved field of density difference between two condensates in segregating atomic binary BECs.
Only recently, the dynamic scaling law has been examined theoretically for domain pattern formations of the BEC systems in different situations \cite{2012Takeuchi, 2013Kudo, 2013Karl, 2014Hofmann},
 and now we are better ready to pioneer the problem of phase-ordering percolation in the quantum fluids.
By performing numerical experiments of segregating BECs, our finite-size-scaling (FSS) analysis revealed that the percolation threshold is $p_c \approx 0.5$ for the initial domain pattern caused by the dynamic instability in strongly segregated BECs, independently of the intercomponent interaction.
 The scaling behavior of SP theory of the initial pattern is sustained until the characteristic domain size grows in the order of the system size.
 We show that an infinite domain wall can exist as an interface between two percolating domains in binary superfluids and found that the wall has a non-integer fractal dimension in a scaling regime during the phase-ordering development with $p=0.5$.

This paper is organized as follows.
In Sec. II, we formulate this system and perform the FSS analysis based on SP theory for the initial domain patterns emerging from the dynamic instability.
 In Sec. III, the scaling behaviors of the percolating domain and an infinite domain wall are described in a unified method independent of the intercomponent interaction.
 It is shown that the scaling behaviors are sustained during the phase-ordering development by rescaling the FSS analysis with the characteristic domain size. 
Section IV is devoted to the conclusion and discussion.

\section{Initial domain patterns}

\subsection{Formulation}

We consider a percolation problem of density domains of quasi-two-dimensional binary BECs in a square box. By introducing the order parameter $\psi_j=\sqrt{n_j} e^{i \theta_j}$ ($j=1,2$) for the $j$th component at low temperatures, the system obeys the Gross--Pitaevskii (GP) Lagrangian \cite{2008Pethick}
\begin{eqnarray}
{\cal L}=\int \( i\hbar \psi_1^* \partial_t \psi_1+i\hbar \psi_2^* \partial_t \psi_2 -K-V \)dxdy 
\label{Lagrangian}
\end{eqnarray}
with the kinetic energy density $K=\sum_j (\hbar^2/2m)|{\bf \nabla}\psi_j |^2$ and the potential energy density
\begin{eqnarray}
V=\frac{g(1+\gamma_{12})}{4}(n_1+n_2)^2+\frac{g(1-\gamma_{12})}{4}(n_1-n_2)^2,
\label{Vpot}
\end{eqnarray}
where $m$ is the particle mass, and $g$ and $g\gamma_{12}$ are the intracomponent and intercomponent coupling constants, respectively.

 The total energy $\int dxdy(K+V)$ and the norm $N_j=\int dx dy n_j$ are conserved quantities during the time development. For $\gamma_{12} >1$, the second term on the right-hand side of \Eq{Vpot} becomes negative with $n_d\equiv n_1-n_2\neq 0$, and binary BECs undergo a spontaneous breaking of the $Z_2$ symmetry with $n_d=n$ or $-n$, leading to domain formation \cite{2008Papp,2009Anderson,2010Tojo,2014Campbell}. The density $n_j$ becomes almost zero in the domains of the $k(\neq j)$th component. A straight domain wall is stabilized under the pressure balance $P_1=P_2$ with the hydrostatic pressure $P_j=gn_j^2/2$ in the domain of the $j$th component. By neglecting the thickness and curvature of the walls, we may write $n_j\approx n \equiv(N_1+N_2)/L^2$ at a place far from domain walls. Without loss of generality, we consider the first component as the percolation element discussed below; then, the occupation rate is defined as $p=N_1/(N_1+N_2)$.

\begin{figure}
\begin{center}
\includegraphics[width=0.98 \linewidth,keepaspectratio]{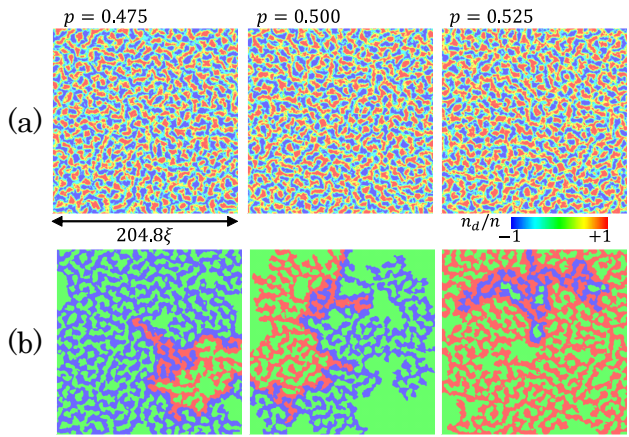}
\end{center}
\caption{(Color online)
(a) The initial domain patterns of the density difference $n_d$ with $l \approx l_0$ for $\gamma_{12}=2.0$, $L=204.8\xi$, and $p=0.475, 0.500, 0.525$.
Domain walls are represented by green regions with lower values of $n_d\sim 0$.
 (b) The maximum domain of (a) is represented by a red (blue) region for the first (second) component.
The maximum domain is defined as a domain with the largest area in each component in the numerical box without taking the periodic boundary condition into account.
} 
\label{MaxDom}
\end{figure}

\subsection{Characteristic scales}

\Figure{MaxDom}(a) shows the characteristic domain patterns emerging during the early stages of the time development for different values of $p$ obtained by numerically solving the GP equations derived from the Lagrangian (\ref{Lagrangian}). We imposed the periodic boundary conditions in our numerical simulations for the sake of computational convenience. The initial state for the numerical computation is the fully miscible state with $n_j(t=0)=N_j/L^2$ with a system size $L$. In our simulation, the dynamic instability is triggered by initially introducing a random fluctuation with a small amplitude $\delta$. For a strong instability, the initial domain pattern emerging in the early stages is characterized by the wave number $k_M$ of the Bogoliubov mode with the maximum imaginary part $\Gamma_M$ in the dispersion \cite{2008Pethick}.

For the case of $(N_1-N_2)/(N_1+N_2)\ll 1$, we have $k_M \sim \sqrt{\gamma_{12}-1}/\xi$ and $\Gamma_M \sim (\gamma_{12}-1)/2\tau$ with $\xi=\hbar/\sqrt{gmn}$ and $\tau=\hbar/gn$. Then, the instability is characterized by the length $l_0$ and time $\tau_0$ as follows:
\begin{eqnarray}
l_0 &\equiv& \frac{\pi \xi}{\sqrt{\gamma_{12}-1}}
\label{l_0}\\
\tau_0 &\equiv& \frac{2\pi\tau}{\gamma_{12}-1} \ln\frac{n}{2\delta^2}
\label{tau_0}
\end{eqnarray}
In fact, we found that the domain patterns become clear around $t=\tau_0$ with the mean interwall distance $l\approx l_0$ [see examples in \Fig{MaxDom}(a)]. Here, $l$ is computed by numerically integrating the total length
\begin{eqnarray}
R=L^2/l
\end{eqnarray}
 of the domain-wall line, defined by $n_d=0$.
Our numerical simulations were done with a square computational mesh. The domain-wall line of $n_d=0$ is defined as a collection of sides between neighboring  meshes with $n_d>0$ and $n_d<0$. A saddle point, an intersection of the lines in the mesh space, occasionally occurs when two domain walls are close to each other. We calculated the average value $\bar{n}_d$ of $n_d$ around the saddle point and then regard the point with $\bar{n}_d>0$ ($<0$) as a point occupied by the first (second) component connecting the diagonal domains with $n_d>0$ ($<0$).
 For $\tau \gtrsim \tau_0$, $l(t)$ obeys a characteristic power law, which is typical behavior in the kink formation dynamics observed in similar systems \cite{2012Takeuchi,2013Kudo,2013Karl,2014Hofmann}; however, this is not the main topic of our work.

\subsection{A relic wall between the maximum domains}

Here, we mention the main target---open domain walls that do not form closed loops. We are interested in long-range open walls, which are distributed across the system. A long-range open wall is realized as the interface between the two domains, both of which simultaneously span the system in the same direction, i.e., from $x=-\infty$ to $\infty$ (or from $y=-\infty$ to $\infty$).
Then, the open wall between the spanning domains is directed to the $x$($y$) direction.
We call such an open domain wall a {\it directed open wall}, and other open walls are referred to as {\it nondirected open walls}.
A directed open wall is deformed into a straight wall owing to some dissipative mechanism, which smooths the domain-wall lines.
To study the dynamics of such an open wall is important since it could survive for long times and thus influences later development of the phase-transition dynamics.
Closed walls shrink and subsequently vanish.
A closed wall can be stabilized around a so-called coreless vortex, in which a component is trapped by the core of a quantized vortex in the other component. The core size of such a vortex is on the order of the thickness of a domain wall, and we found that not so many vortices appear in our system. Thus, the effect is not important to our analysis.

A long-range open wall exists as an interface between the two domains, whose area is the largest in each component. In our dynamical system, evaluation of the survival of open domain walls is not straightforward. A directed open wall can be transformed into a nondirected open wall after a few reconnections of the wall with other short walls. An example of such a nondirected wall is demonstrated in \Fig{MaxDom}(b) for $p=0.500$, where the wall sandwiched between the maximum domains runs from $x=-L/2$ to $y=L/2$. Conversely, it is easy to create a directed wall from such a long nondirected wall. Thus, an open wall sandwiched between the maximum domains has the potential to survive as ``a relic'' of the transition. We call such a long-range wall the {\it relic wall}. We will evaluate the possibility of a relic wall in connection with the statistical properties of the maximum domains. SP theory \cite{1994Stauffer} is useful to the evaluation.

\subsection{Percolation probability}

To demonstrate the direct relation of our system with the percolation problem, we computed the ensemble average of the area $S_M(p,L)$ of the maximum domain.
In SP theory [20], a percolation problem is analyzed by calculating the area of the maximum cluster divided by the system area (see also the Appendix). Then, our
 percolation problem is analyzed by the quantity
\begin{eqnarray}
P(p,L)\equiv \<S_M(p,L)\>_l/L^2,
\label{PSM}
\end{eqnarray}
where $\<\cdots\>_l$ represents the ensemble average with $l$ fixed.
We took the ensemble average of all the domain patterns satisfying the condition $|l(t)-l_0|/l_0 \lesssim 0.005$ during each development by performing numerical simulations $N_L$ times with $N_L\geq 6553.6\times \xi/L$.
 We found that the maximum domain spans the system probably if $p \gtrsim 0.5$ ($p \lesssim 0.5$) for the first (second) component [see \Fig{MaxDom}(b)]. Actually, $P(p,L)$ increases rapidly from $p\sim 0.5$. This upturn becomes sharper for a larger system size, and the probability asymptotically approaches $P(p,L)\to p$ for $p\to 1$. These behaviors are consistent with the finite-size effect in the SP systems \cite{1994Stauffer}.

\begin{figure}
\begin{center}
\includegraphics[width=0.98 \linewidth,keepaspectratio]{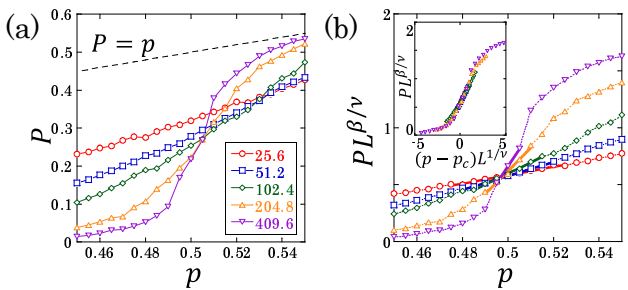}
\end{center}
\caption{(Color online) 
(a) Plots of the probability $P(p,L)$ for $\gamma_{12}=2.0$ and $L/\xi=25.6, 51.2, 102.4, 204.8, 409.6$. (b) Plots of the parameter $P(p,L)L^{\beta/\nu}$. The threshold $p_c$ is determined by averaging the positions of all intersection points between the fitting functions represented with bold solid lines. The inset is the finite-size-scaling plot of (a).
} 
\label{p_P}
\end{figure}

To confirm the scaling behavior of SP in our system, we perform a FSS analysis \cite{1994Stauffer}.
It is convenient to utilize the property of the parameter $P(p,L) L^{\beta/\nu}$ with the critical exponents $\beta=5/36$ and $\nu=4/3$ \cite{1994Stauffer}. This parameter takes a universal value independent of the system size $L$ at $p=p_c$:
\begin{eqnarray}
P(p_c,L)L^{\beta/\nu}=const.
\label{PLconst}
\end{eqnarray}
This property is used to estimate the percolation threshold \cite{2012Chandra}. This estimation is sufficient for our main purpose, although the so-called Binder parameter is typically used to determine the critical point of the phase transition \cite{Binder1981}. \Figure{p_P} shows the $p$ dependence of $P(p,L)L^{\beta/\nu}$ for several values of $L$. The plots with different values of $L$ collapse onto a point at $p\sim 0.5$.

\subsection{Percolation threshold}

The percolation threshold is estimated by fitting a plot with a linear function in the least-squares method for $0.5-\delta p_L \leq p \leq 0.5+\delta p_L$ with $\delta p_L=0.005 \times 409.6 \xi/L$ [see in \Fig{p_P}(b)]. We averaged the positions of all intersection points between the fitting lines with two different values of $L$ and obtained $p_c=0.497(2)$ for $\gamma_{12}=2.0$.
There are intrinsically large fluctuations around the threshold $p_c\approx 0.5$. Thus, to achieve the threshold value with satisfactory accuracy, we carried out many more simulations for $0.5-\delta p_L \leq p \leq 0.5+\delta p_L$. We computed more than $N_L\times 4$ times for $0.485\leq p \leq 0.515$ and $N_L\times 8$ times for $p=0.5$.

 According to the FSS hypothesis \cite{1994Stauffer}, the quantity $P(p,L)$ is scaled by a scaling function $F(x)$ of $x=L^{1/\nu}\Delta p$ with $\Delta p=p-p_c$ as
\begin{eqnarray}
P(p,L)L^{\beta/\nu}=F(L^{1/\nu}\Delta p).
\end{eqnarray} 
The scaling plot is successful when using the above value of $p_c$ [see inset of \Fig{p_P}(b)]. This result clearly shows that our system actually obeys SP theory.

To gain a deeper insight, it is fruitful to investigate how the percolation behavior depends on the intercomponent interaction, which affects the correlation of the domain patterns. It is known that the percolation threshold is sensitive to the correlation between elements, i.e., the number of vertices in bond percolation and the intersite interaction in annealed percolation \cite{1994Stauffer,1978Duckers,1975Odagaki,1993Wollman}.
The most important report related to us is Ref. \cite{1997Nolte}, where the authors investigated the probability of the simultaneous percolation of the two components in the two-dimensional model of two-phase fluids and showed the percolation threshold can change depending on the correlation.

In our system, namely, a two-phase superflow, the correlation is controllable via the intercomponent interaction: the coherence (or healing) length $\xi_d\sim \xi/\sqrt{\gamma_{12}-1}$ of the field $n_d$. However, we found that the percolation threshold remains $p_c \approx 0.5$ insensitively to $\gamma_{12}$; $p_c=0.498(4), 0.498(1)$, and $0.496(2)$ for $\gamma_{12}=1.7, 2.5$, and $3.0$, respectively. The point is that segregated binary BECs are classified into strong segregation ($\gamma_{12} \gtrsim 1.7$) and weak segregation ($1<\gamma_{12} \lesssim 1.7$), which are related to the competition between $\xi_d$ and the healing length of the total density $n_1+n_2$ \cite{1998Ao}. Nevertheless, we did not find any qualitative change for $\gamma_{12} \geq 1.7$.

It is believed that the percolation threshold of a symmetric random scalar field is $p_c=0.5$ in a two-dimensional continuum according to the intuitive argument in Ref. \cite{Zallen1971}.
Our system is regarded as a continuum system since the numerical simulations were done with a square computational mesh whose size is small enough compared to any relevant length scale of this system.
Supposing the field $n_d$ plays the role of the random scalar field, our result seems to be consistent with the argument of Ref. \cite{Zallen1971}.

\section{Rescaling analysis}

\subsection{Rescaling plot}

The fact that the percolation property is not sensitive to the intercomponent interaction makes us expect that domain patterns with different $\gamma_{12}$ can be universally described as having the same statistical property. The FSS analysis is applied after $L$ is scaled by the characteristic domain size, that is the mean interwall distance $l=l_0(\gamma_{12})$. Then, the probability in \Eq{PSM} is rescaled by the effective system size $\tilde{L}=L/l$ as
\begin{eqnarray}
\tilde{P}\( p, \tilde{L}\) \tilde{L}^{\beta/\nu}=\tilde{F}\( \tilde{L}^{1/\nu}\Delta p \),
\label{Peff}
\end{eqnarray}
where we used the relation $\tilde{P}(p,\tilde{L})=P(p,L)$ and the rescaling function $\tilde{F}$.

\begin{figure}
\begin{center}
\includegraphics[width=0.98 \linewidth,keepaspectratio]{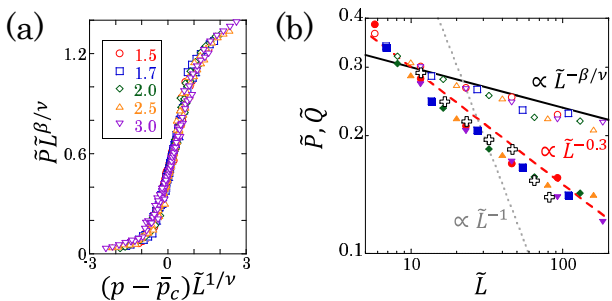}
\end{center}
\caption{(Color online) 
(a) The rescaling plot with the threshold $\bar{p}_c$. The probability $P(p,L)$ with $L/\xi=25.6, 51.2, 102.4, 204.8, 409.6$ is rescaled for $\gamma_{12}=1.5, 1.7, 2.0, 2.5, 3.0$.
(b) The $\tilde{L}$ dependence of the relic probability $\tilde{Q}$ (filled marks).
Shapes of the filled marks corresponds to those of the legend symbol in (a)  for reference of the values of $\gamma_{12}$.
 A plot of the rescaled percolation probability $\tilde{P}$ at $p=0.5$ is displayed with the same marks as in (a). White crosses represent the dynamical plot of $\tilde{Q}(\tilde{L}(t))$ with $\gamma_{12}=2.0$ and $L/\xi=204.8$ for several values of $\tilde{L}(t)=L/l(t)$.
} 
\label{Rescaling}
\end{figure}

This conjecture is confirmed by the successful overlapping of the rescaling plots by using the average value of the thresholds of $\gamma_{12}=1.7, 2.0, 2.5$, and $3.0$ and $\bar{p}_c \approx 0.497$, in \Fig{Rescaling}(a).
Note that the numerical experiments become more difficult to perform with decreasing $\gamma_{12}$ because the computation time and the required system size are proportional to $\tau_0$ and $l_0$, respectively.
 For this reason, we could not determine the percolation threshold for $\gamma_{12}=1.5$ with satisfactory accuracy.
 However, our rescaling plot successfully includes the data for $\gamma_{12}=1.5$.
 From this fact, we conclude that the percolation properties of this system are nearly independent of $\gamma_{12}$, although we cannot examine the limit $\gamma_{12} \to 1$ because of the difficulty in numerical simulation with $\tau_0 \to \infty$ and $l_0 \to \infty$.

\subsection{Relic probability}

Now, let us return to the relic-wall problem. To evaluate the spatial distribution of the relic wall, we introduce the relic-wall length $R_M$ and the probability that a specific wall segment belongs to the relic wall:
\begin{eqnarray}
Q(L)\equiv \<R_M/R\>_l=l\<R_M\>_l/L^2.
\label{Q}
\end{eqnarray}
We consider the rate $p=0.5$, with which the dynamical analysis is valid, as discussed later. Because this rate is very close to the percolation threshold, we assume that the two maximum domains have the scaling behavior described by SP theory. From Eqs. (\ref{PSM}) and (\ref{PLconst}), we have
\begin{eqnarray}
S_M(p \to p_c, L) \propto L^{D_S},
\end{eqnarray}
 and the percolating domain has the fractal dimension
\begin{eqnarray}
D_S=2-\beta/\nu.
\end{eqnarray}
Similarly, we assume the scaling behavior
\begin{eqnarray}
R_M(L)\propto L^{D_R}
\label{RML}
\end{eqnarray}
with the fractal dimension $D_R$ of the relic wall.

In general, a fractal dimension $D$ of an object in two-dimensional space takes values ranging from zero to two depending on the scale of view. For example, in our system, a closed domain wall with a finite size has $D=0$ if it is viewed on the scale of $L \to \infty$. Similarly, if the domain wall forms a straight line across the system, we have $D=1$. When a single domain wall is distributed uniformly throughout the system, its fractal dimension is $D=2$.

Because of the negligible deviation of the percolation threshold $p_c$ from the critical rate ($p=0.5$) of the phase-ordering kinetics, the fractal dimension $D_R$ is restricted by the scaling behavior of the percolating domains.
 If domain walls are homogeneously distributed throughout the system, the relic probability $Q(L)$ is considered to be the ratio of the area occupied by the relic wall to the system area $L^2$.
Because the relic wall is sandwiched between percolating domains, the area ratio must be smaller than the right-hand side of \Eq{PSM}, which is the area ratio of the maximum domains to $L^2$. Then, for large $L$, the following condition should be satisfied:
\begin{eqnarray}
1 \leq D_R \leq D_S.
\label{DRDS}
\end{eqnarray}
The lowest limit represents the requirement $R_M \gtrsim L$, coming from the fact that the length of a relic wall between two percolating domains is not smaller than the linear size of the system.

A rescaling analysis is convenient to produce a unified description of the scaling behavior of our system. From the fractal behavior of the maximum domain, we have
\begin{eqnarray}
\tilde{P}(p\to p_c, \tilde{L}) \sim \tilde{L}^{-\beta/\nu}.
\label{PeffLeff}
\end{eqnarray}
In the same way, the relic probability $\tilde{Q}(\tilde{L})=Q(L)$ is written as
\begin{eqnarray}
\tilde{Q}(\tilde{L})\sim \tilde{L}^{-h}
\label{Qeff}
\end{eqnarray}
with the exponent
\begin{eqnarray}
h\equiv 2-D_R.
\end{eqnarray}
Then, the condition (\ref{DRDS}) reduces to
\begin{eqnarray}
\beta/\nu \leq h \leq 1.
\label{betah}
\end{eqnarray}
If the relic wall does not exhibit any scaling behavior, $D_R$ should be an integer, that is, $h=1$  from the restriction (\ref{DRDS}) with $D_R=1$.

\Figure{Rescaling}(b) shows the rescaling plots of $\tilde{Q}(\tilde{L})$ for the initial domain pattern with $\tilde{L}=L/l_0(\gamma_{12})$. The condition (\ref{betah}) is actually satisfied with $h\approx 0.3$. For comparison, we plot $\tilde{P}(p=0.5,\tilde{L})$, which is consistent with \Eq{PeffLeff}. Our scaling analysis is well applicable for $L\gg l$. This is why $\tilde{P}$ does not obey this law below $\tilde{L} \sim 10$. Our results show that a relic wall has scaling behavior with a noninteger fractal dimension with $D_R \approx 1.7$.

\subsection{Dynamic finite-size-scaling analysis}

The relic wall will maintain the scaling behavior for a long time during the phase-ordering development because the system with $p\sim 0.5$ obeys the dynamic scaling law \cite{Bray1994,2014Hofmann}. The law states that the domain patterns at a later time are statistically similar to those at an early time if the patterns are scaled by the characteristic domain size $l(t)$. Then, we may replace $\tilde{L}=L/l_0$ in the above analysis with the dynamical value $\tilde{L}(t)=L/l(t)$.

\begin{figure}
\begin{center}
\includegraphics[width=0.98 \linewidth,keepaspectratio]{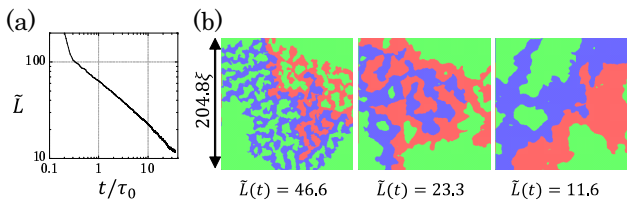}
\end{center}
\caption{(Color online) 
(a) A typical time evolution of $\tilde{L}(t)$ and (b) the snapshots of the maximum domains in a numerical simulation for $p=0.5$ at $t/\tau_0 \approx 2, 9$, and $33$ with $\tilde{L}=46.6, 23.3$, and $11.6$, respectively.
(b) The maximum domains are represented by a red (blue) region for the first (second) component in the same manner as Fig. \ref{MaxDom} (b).
} 
\label{RelicWall}
\end{figure}

To test this argument, we computed the relic probability $\tilde{Q}(\tilde{L}(t))$ for $l(t)/l_0=0.8, 1.0, 1.4, 2.0, 2.8, 3.9$,  and $5.6$ by averaging many numerical experiments for $\gamma_{12}=2.0$ and $L/\xi=204.8$ [see \Fig{Rescaling}(b) and \Fig{RelicWall}]. We numerically confirmed that the relic wall maintains its scaling behavior with $h \approx 0.3$ during the phase-ordering development until a later time with $\tilde{L}(t) \sim 10$.

 It is noteworthy that an infinite domain wall maintains its scaling behavior in our system; in contrast,
 the system of infinite cosmic strings is believed to not be scale invariant in numerical cosmology \cite{1984Vachaspati,2000Vachaspati}.
 There, the fractal dimension of the infinite cosmic string was estimated to be $2$ approximately, connected with that of the random walk.
 In this sense, it is also interesting that the fractal dimension $D_R$ is between those of the random walk and the self-avoiding walk, $2$ and $4/3$ \cite{Madras2013}, respectively.

Finally, it should be mentioned that the fractal dimension $D_R$ of the relic wall in our system is larger than that ($\lesssim 1.6$) of the longest disclination line in a quenched nematic liquid crystal \cite{1987Orihara},
 where the cell of twisted liquid crystal is so thin that the system can be regarded as a two-dimensional system of a non-conserved field and the disclination as a one-dimensional domain wall \cite{1986Orihara}.
 This discrepancy may be attributable to the two facts that the temperature in the cell gradually decreased during the relaxation dynamics and the system size was not large enough to neglect the fluctuation and to confirm clearly the scaling behavior.

\section{Conclusion and discussion}
In conclusion, we found that the percolation threshold of segregating binary BECs is approximately 0.5. An infinite domain wall with a noninteger fractal dimension can survive in the dynamic scaling regime, where the characteristic domain size ($l$) is much larger than the system size ($L$).
Experimentally, we will observe a relic domain wall in an atomic cloud whose size is larger than at least dozens of times the length $\xi$ for $\gamma_{12}=2.0$.
 It is expected that our scaling analysis is generally applicable to two-dimensional systems for both conserved and nonconserved fields.
In the last stage with $l \lesssim L$, there appears a remarkable difference between the two fields.
 A large closed wall or a long-range open wall is left in the final state for conserved fields since $p$ is fixed to $0.5$, while the stripe states or the ground state with $p=0$ or $1$ are realized for nonconserved field \cite{Arenzon2007}.

 It is a nontrivial problem whether an infinite domain wall exhibits a noninteger-fractal behavior even when $p_c$ is not so close to $0.5$.
 It must be instructive to investigate phase-ordering percolation in three dimensions,
 where the percolation threshold becomes smaller than that in two dimensions in general \cite{1994Stauffer}.
The application of our analysis to different phase-ordering systems would also be helpful to verify the exponent $h$ and its relation with other critical exponents.

\begin{acknowledgments}
We are grateful to H. Orihara, T. Hiramatsu, M. Tsukamoto, H. Ishihara, T. Odagaki, M. Takahashi, and M. Kobayashi for useful discussions and comments. This work was partly supported by KAKENHI from the Japan Society for the Promotion of Science (Grants No. 25887042 and No. 26870500). This work was also partly supported by the Topological Quantum Phenomena (Grants-in-Aid No. 22103003) for Scientific Research on Innovative Areas from the Ministry of Education, Culture, Sports, Science and Technology of Japan.
\end{acknowledgments}

\begin{appendix}
\section{Critical behavior close to the percolation threshold}
Here, we make a brief introduction of the critical behavior of the maximum cluster described by SP theory.
The percolation probability is the probability that a given point belongs to an infinite cluster.
In the limit of infinite system size $L\to\infty$, the probability is written as
\begin{eqnarray}
P(p)=\lim_{L\to \infty} \frac{S_M(p,L)}{L^2},
\label{P_p}
\end{eqnarray}
where $p$ and $S_M$ are the occupation rate and the area of the maximum cluster of the element considered.

According to the percolation theory, we have the following critical behaviors around the critical point $p_c$,
\begin{eqnarray}
&&P(p)\sim (p-p_c)^\beta~~~(p \geq p_c),
\label{A_P_pc}\\
&&\xi_P(p)\sim |p-p_c|^{-\nu}.
\label{A_xi_pc}
\end{eqnarray}
Here, $\xi_P$ is the correlation length between two points in the same cluster. These equations reduces to
\begin{eqnarray}
P(p) \sim \xi_P^{-\beta/\nu}~~~(p\geq p_c).
\label{A_P_pc}
\end{eqnarray}

The finite-size-scaling hypothesis states that the statistical behavior in a system with a linear size $L$ must be characterized by the ratio $L/\xi_P$. 
Correspondingly, we shall evaluate a quantity
\begin{eqnarray}
P(p,L)\equiv \frac{S_M(p,L)}{L^2},
\label{A_P_pL}
\end{eqnarray}
which exhibits the finite size effect through a scaling function $f$ around the critical point as
\begin{eqnarray}
P(p,L) \sim  \xi_P^{-\beta/\nu} f(L/\xi_P).
\label{A_P_f}
\end{eqnarray}
From this formula, we suppose a scaling law
\begin{eqnarray}
P(p,L)L^{\beta/\nu}=F(L^{1/\nu}\Delta p)
\label{A_Slaw}
\end{eqnarray}
with $\Delta p=p-p_c$.

The system size $L$ is only a relevant length scale at $p=p_c$ since $\xi_P$ diverges as $\propto |p-p_c|^{-\nu}$ with $\nu>0$.
Thus, $P(p_c,L)$ should be a function of not $\xi_P$ but $L$, and we have
\begin{eqnarray}
f(L/\xi_P)\sim(L/\xi_P)^{-\beta/\nu}~~~(p\to p_c).
\label{A_f_pc}
\end{eqnarray}
By substituting \Eq{A_P_f} with the above formula into \Eq{A_P_pL}, we have
\begin{eqnarray}
S_M(p \to p_c,L)\sim L^{D_S},~~~D_S=2-\beta/\nu
\label{A_SM_D}
\end{eqnarray}
Here, this equation shows that the maximum cluster at the critical point has a fractal dimension $D_S$.

To derive the formula of \Eq{PLconst}, we introduce proportional constants $C_P$ and $C_f$ as $P(p,L) \to C_P\xi_P^{-\beta/\nu} f(L/\xi_P)$ and $f(L/\xi_P)\to C_f(L/\xi_P)^{-\beta/\nu}$ for $p \to p_c$.
From these relations one obtains
\begin{eqnarray}
P(p_c,L)L^{\beta/\nu}=const.
\label{A_PLconst}
\end{eqnarray}
This shows that the quantity $P(p,L)L^{\beta/\nu}$ should approach a universal value at $p = p_c$ independent of $L$.
This argument is consistent with the scaling law (\ref{A_Slaw}) with $\Delta p=0$.

\end{appendix}


\end{document}